\documentclass[prb,twocolumn,showpacs]{revtex4}

\usepackage{graphicx}
\usepackage{dcolumn}
\usepackage{bm}
\usepackage{amsmath}

\begin{document}

\title{Non-Markovian qubit dynamics in the presence of 1/f noise}

\author{Guido Burkard}
\affiliation{Institute of Theoretical Physics C, 
RWTH Aachen University, D-52056 Aachen, Germany}
\affiliation{Department of Physics, University of Konstanz, D-78457
  Konstanz, Germany}

\begin{abstract}
Within the lowest-order Born approximation, we calculate the exact dynamics 
of a qubit in the presence of 1/f noise, without 
Markov approximation.  We show that the non-Markovian qubit
time-evolution exhibits asymmetries and beatings that can be observed 
experimentally and cannot be explained within a Markovian theory.
The present theory for 1/f noise is relevant for both spin- and superconducting
qubit realizations in solid-state devices, where 1/f noise is ubiquitous.
\end{abstract}




\maketitle

\section{Introduction}
Random telegraph noise
has been encountered in a wide range of situations
in many different areas of physics \cite{DH81}.
A typical example in condensed matter physics is that of a resistor
coupled to an ensemble of randomly switching impurities, producing
voltage fluctuations with a spectral density that scales inversely 
proportional with the frequency, hence the name ``1/f noise''.
The quest to build and coherently control quantum two-level systems 
functioning as qubits
in various solid state systems has once more highlighted the importance of 
understanding
1/f noise, being a limitation to the quantum coherence 
of such devices.  

\begin{figure}
\includegraphics[width=8cm]{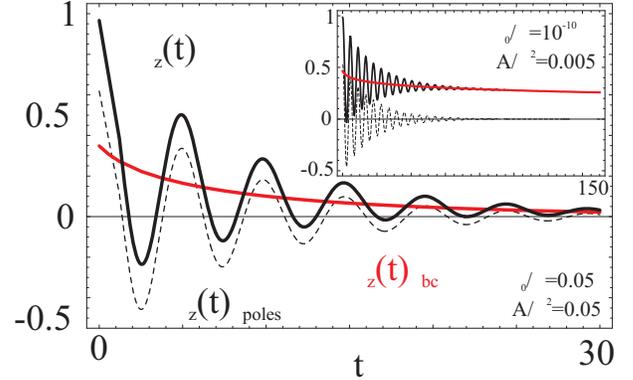}
\caption{(Color online) Non-Markovian time-evolution of the unbiased ($\epsilon=0$) qubit (spin) 
$z$-component $\langle \sigma_z(t)\rangle$, for $A/\Delta^2=0.05$ and 
$\gamma_0/\Delta=0.05$ (solid black line).
The Markovian pole contribution $\langle \sigma_z(t)\rangle_{\rm poles}$
is plotted as a dashed line for comparison. 
The essential non-Markovian part is non-exponential and given by the 
branch cut contribution $\langle \sigma_z(t)\rangle_{\rm bc}$ (red solid line).
Inset:  Plot for $A/\Delta^2=0.005$ and $\gamma_0/\Delta=10^{-10}$. Here,
the essential non-Markovian part is the long-time asymmetry which carries
information about the initial state.
}
\label{fig:plot}
\end{figure}

The description of low-frequency noise (such as 1/f noise) is
complicated by the presence of long-time correlations
in the fluctuating environment which prohibit the use of the Markov
approximation.   Only in few cases, 
non-Markovian effects have been taken into account exactly, e.g., for 
the relaxation of an atom to thermal equlibrium \cite{DK}.
Here, we are interested in the decoherence and relaxation of a qubit, 
i.e., a single two-level system (spin 1/2).  For the spin-boson model, i.e., a qubit 
coupled to a bath of harmonic oscillators, the dynamics has been 
calculated within a rigorous Born approximation without making a
Markov approximation \cite{LD03,DL05}.  Here, we carry out a similar analysis
for 1/f noise and find even stronger effects than in the spin-boson
case (see Fig.~\ref{fig:plot}).
 
Charge and to some extent (via the spin-orbit interaction) spin qubits
in quantum dots \cite{LD98} formed in semiconductor \cite{Jung04} or
carbon \cite{Tobias08} structures are subject to 1/f noise.
In superconducting (SC) Josephson junctions, SC interference
devices (SQUIDs), and SC qubits, 1/f noise has been extensively studied 
experimentally 
\cite{Simmonds04,Wellstood04,Astafiev04,Ithier05,Mueck05,Eroms06,Astafiev06,Yoshihara06} 
and theoretically \cite{vanHarlingen04,Koch07}.

Even where the origin of 1/f noise is known, the induced decoherence
is not fully understood.
Most theoretical work is either restricted to longitudinal fluctuations
or employs a Markov approximation.  Here, we present a 
calculation of the qubit dynamics in the presence of 1/f noise which is exact 
within the lowest-order Born approximation.  In particular, we make no use of
a Markov approximation.  In contrast to earlier calculations
\cite{Martinis03,MS04,Rabenstein04,Schriefl06}, we allow for
arbitrary qubit Hamiltonians and include
transverse as well as longitudinal (phase) 1/f noise.
Non-Gaussian 1/f noise originating from
few fluctuators was studied in \cite{Paladino02,Galperin06,Bergli06a}, 
while numerical
studies using an adiabatic approximation were carried out in \cite{Falci05}.
The coupling to a single fluctuator was also studied \cite{Saira07}.

\section{Model}
We model the qubit (spin 1/2) coupled to
a bath of two-level fluctuators with the Hamiltonian 
\begin{equation}
H=H_S+H_B+H_{SB}
\end{equation}
with
\begin{eqnarray}
  \label{eq:HS}
  H_S &=& \Delta \sigma_x + \epsilon \sigma_z, \\
  H_{SB} &=& \sigma_z X,
\end{eqnarray}
where $\sigma_x$ and $\sigma_z$ are Pauli matrices describing the
qubit and $X= \sum_{i=1}^N v_i \sigma_z^i$ where
$\sigma_z^i$ operates on the $i$-th fluctuator.
In a SC qubit, $\Delta$ and $\epsilon$ denote the tunneling  and
energy bias between the two qubit states.  In a spin qubit,
$\epsilon$ is the Zeeman splitting and $\Delta$ a transverse field.
The bath Hamiltonian $H_B$ need not be provided explicitly;
it is sufficient to know the auto-correlator
$C(t)= \langle X(0) X(t)\rangle$
of the bath operator $X(t)$,
where $\langle\ldots\rangle={\rm Tr}_B(\ldots\rho_B)$ denotes
a trace over the bath degrees of freedom with the bath density 
matrix $\rho_B$. We can further assume that the fluctuators are
unbiased, $\langle X(t)\rangle =0$.
For independent two-level fluctuators with switching
rates $\gamma_i$, one obtains
\begin{equation}
\label{eq:Ct}
C(t) = \sum_i v_i^2  \langle \sigma_i(t) \sigma_i(0)\rangle
      =   \sum_i v_i^2 e^{-\gamma_i |t|}.
\end{equation}
The noise spectral density is the Fourier transform
\begin{equation} \label{FT}
S(\omega) = \int_{-\infty}^\infty \! dt \, C(t) e^{-i\omega t}
= \sum_i (2 v_i^2 \gamma_i)/(\gamma_i^2+\omega^2).
\end{equation}
While this correlator describes essentially classical bath dynamics
(as is commonly assumed for 1/f noise),
it should be emphasized that our model is \textit{not} classical, because
the $[H_{SB},H_S]\neq 0$.
In the case of a large number of fluctuators, the sum in $C(t)$
can be converted into
an integral.  For 1/f noise, one typically assumes a distribution of 
fluctuators of the form $P(v,\gamma) \propto 1/\gamma v^\beta$,
where both $v$ and $\gamma$ are limited by upper and lower cut-offs
\footnote{We assume $\beta >3$ to ensure that a large number of 
fluctuators over the entire range of $v$ couples to the qubit and  
the assumption of Gaussian noise is justified.}.
The spectral density of the ensemble of fluctuators
then becomes
\begin{equation}
  \label{eq:Sdef}
  S(\omega) \propto \int_{v_{\rm min}}^{v_{\rm max}}
\int_{\gamma_0}^{\gamma_c} dv\, d\gamma P(v,\gamma) 
\frac{2 v^2 \gamma}{\gamma^2+\omega^2}.
\end{equation}
For $\gamma_0=0$ this yields 1/f noise of the form 
$S(\omega) \propto 1/|\omega|$.
The divergence at low frequencies is cut off
by the finite duration of a qubit measurement, if not by other
effects at even shorter times.
A low-frequency cut-off $\gamma_0>0$ yields
\begin{equation}
  \label{eq:S1}
  S(\omega)=2\pi A \frac{\arctan(\omega/\gamma_0)}{\pi}\frac{1}{\omega},
\end{equation}
where $A$ depends on the cut-offs and the exponent $\beta$.
For $\gamma_0\rightarrow 0$, we recover 
$S(\omega)\rightarrow 2\pi A/|\omega|$.
Inverting the above Fourier transform, we obtain
\begin{equation}
C(t)= -A\, {\rm Ei}(-\gamma_0 |t|),
\end{equation}
where ${\rm Ei}$ denotes the exponential integral function.

\section{Qubit dynamics}
The density matrix $\rho$ of the total system, consisting of the
qubit and the bath, obeys the Liouville equation,
$\dot\rho(t) = -i[H,\rho(t)]$.
The time evolution of the reduced density matrix
of the qubit alone
$\rho_S(t) = {\rm Tr}_B \rho$
is then determined by the generalized master equation (GME)
\cite{LD03,DL05}
\begin{equation}
  \label{eq:GME}
  \dot \rho_S(t) = -i [H_S,\rho_S(t)] -i\int_0^t \Sigma(t-t') \rho_S(t') dt' ,
\end{equation}
where  the self-energy superoperator $\Sigma(t)$ gives rise
to memory effects, i.e., the time evolution of $\rho_S(t)$
depends on the state $\rho_S(t')$ at all earlier times $t'\le t$.
Therefore, the qubit dynamics is inherently non-Markovian.
Expanding the right-hand side of the GME in orders
of $H_{SB}$ and only keeping
the lowest (second) order, one obtains $\Sigma$ 
in (lowest-order) Born approximation
$\Sigma(t)\rho_S =
-i{\rm Tr}_B [H_{SB},e^{-itH_0}[H_{SB},\rho_S\otimes \rho_B]e^{it H_0}]$,
where $H_0=H_S+H_B$.

Introducing the Bloch vector
$\langle {\bm \sigma}(t)\rangle = {\rm Tr}_S {\bm \sigma}\rho_S(t)$,
where $ {\bm \sigma}=(\sigma_x,\sigma_y,\sigma_z)$ is a vector
of Pauli operators,
we write the GME as a generalized Bloch equation
\begin{equation}
  \label{eq:GBE}
  \langle \dot {\bm \sigma}\rangle = R * \langle {\bm \sigma}\rangle + {\bm k},
\end{equation}
where the star denotes convolution and \cite{LD03,DL05}
\begin{equation}
  \label{eq:Rmatrix}
  R(t)=
\begin{pmatrix}
-\frac{E^2}{\Delta^2}\Gamma_1(t) & -\epsilon\delta(t) + \frac{E}{\Delta}K_y^+(t) & 0\\
\epsilon \delta(t)  -\frac{E}{\Delta}K_y^+(t)& -\Gamma_y(t) & -\Delta\delta(t)\\
0 & \Delta\delta(t) & 0
\end{pmatrix}
\end{equation}
with $E=\sqrt{\Delta^2 +\epsilon^2}$ and \cite{LD03,DL05}
$\Gamma_1(t) = (2 \Delta/E)^2\cos(Et)C'(t)$,
$\Gamma_y(t) = (2\Delta/E)^2(1+(\epsilon/\Delta)^2\cos(E t))C'(t)$,
and $K_y^+(t) = (4 \epsilon \Delta/E^2)\sin(Et)C'(t)$,
where $C'(t)$ and $C''(t)$ denote the real and imaginary parts of $C(t)$.
Since for 1/f noise, $C''(t)=0$, we find ${\bf k}(t)=0$ \cite{LD03,DL05}.
As shown in \cite{LD03,DL05}, Eq.(\ref{eq:GBE}) can be solved
by means of the Laplace transform (LT)
$f(s) = \int_0^\infty f(t) e^{-ts}\,dt$, where
\begin{equation}
  \label{eq:GBE-L}
  \langle {\bm \sigma}(s)\rangle = \left(s-R(s)\right)^{-1} 
    \left( \langle {\bm \sigma (t=0)\rangle} - {\bm k}(s)\right).
\end{equation}
The LT $R(s)$ of $R(t)$, has entries
according to Eq.~(\ref{eq:Rmatrix}),
with $\delta(t)$ replaced by $1$, and,
for 1/f noise
\begin{eqnarray}
\Gamma_1(s) &=& (2A/E^2) \Delta ^2 \left(C(s+iE)+C(s-iE)\right),\\
\Gamma_y(s) &=& (2 A/E^2) \left(2 \Delta ^2C(s) \right.\nonumber\\
  & & \left.  +\epsilon ^2 \left(C(s+iE)+C(s-iE)\right)\right), \\
K_y^+(s) &=& i(2A/E^2)  \Delta  \epsilon  \left(C(s+iE)-C(s-iE)\right),\quad
\end{eqnarray}
where the LT of the correlator $C(t)$ in Eq.~(\ref{eq:Ct}) is
\begin{equation}
\label{eq:Cs}
C(s)=\frac{A}{s}\log\left(1+s/\gamma_0\right).
\end{equation}
We recover $\langle{\bm \sigma}(t)\rangle$ from  
$\langle{\bm \sigma}(s)\rangle$ by way of an
inverse LT as carried out below,
first for the special case of an unbiased qubit ($\epsilon=0$) 
and then for the general case.

\section{Unbiased qubit}
\begin{figure}[b]
\includegraphics[width=8cm]{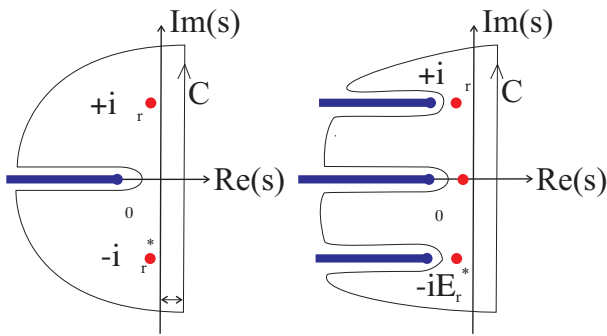}
\caption{(Color online) Analytic structure of $\langle\sigma_z(s)\rangle$ in the 
complex $s$ plane, for (a) the unbiased case,
$\epsilon=0$ and (b) the biased case, $\epsilon\neq 0$.  
Red dots denote poles, blue lines branch cuts.
}
\label{fig:analytic}
\end{figure}
We first assume that the qubit is prepared at time $t=0$ in one of the eigenstates 
$|0\rangle=|\!\!\uparrow\rangle$ of $\sigma_z$, i.e., 
$\langle {\bm \sigma}\rangle=(0,0,1)$, 
and that the qubit is unbiased, $\epsilon=0$.
If the fluctuators were absent the qubit would undergo a
precession about the $x$ axis, $\langle \sigma_z(t)\rangle=\cos(\Delta t)$.
Due to the presence of the fluctuators, we find 
(see also Appendix \ref{app:propagator})
\begin{equation}
  \label{eq:sol-trans}
  \langle \sigma_z  (s)\rangle
   = \frac{s^2 + 4 A \log(1+s/\gamma_0)}
          {s\left(s^2+\Delta^2+4 A \log(1+s/\gamma_0)\right)}.
\end{equation}
We expand $\langle\sigma_z(s)\rangle$ in leading order of $A$,
\begin{equation}
  \label{eq:sol-trans-weak}
  \langle\sigma_z(s)\rangle = \frac{s}{s^2+\Delta^2}+4A\Delta^2 \frac{\log(1+s/\gamma_0)}{s(s^2+\Delta^2)^2} +O(A^2).
\end{equation}
The coherent spin oscillations in the time domain are
obtained from the inverse LT, the so-called
Bromwich integral \cite{LD03,DL05} (see Fig.~\ref{fig:analytic}),
$\langle\sigma_z(t)\rangle 
= \frac{1}{2\pi i}\lim_{\eta\downarrow 0}\int_{-i\infty+\eta}^{i\infty+\eta}\langle\sigma_z(s)\rangle e^{ts}\, ds$.
The integral contour can be closed in the left complex half-plane
${\rm Re}(s)<0$ (Fig.~\ref{fig:analytic}).
The behavior of $\langle\sigma_z(t)\rangle$ is therefore given by the 
analytic structure of $\langle\sigma_z(s)\rangle$ in the left half-plane,
see Fig.~\ref{fig:analytic}.
In the absence of the fluctuating environment ($A=0$), 
$\langle\sigma_z(s)\rangle$ has two poles at $s=\pm i\Delta$ which 
yield $\langle\sigma_z(t)\rangle=\cos(\Delta t)$,
as expected.  The coupling to the environment has two effects:
(i) a shift of the poles, and (ii) the appearance of a branch point (bp) due to the
logarithm in Eq.~(\ref{eq:sol-trans-weak}) and the associated branch
cut (bc)
that we choose to lie on the real axis between $-\gamma_0$ and $-\infty$.
Here, it should be noted that in the case of an unbiased qubit,
the presence of 1/f noise does not lead to the appearance of 
a pole on the real axis, and thus there is only pure dephasing and
no $T_1$ type decay 
(spin relaxation), in contrast to other types of environment \cite{DL05}.
The exact shift of the poles has been calculated numerically from
Eq.~(\ref{eq:sol-trans}).
To lowest order in $A$, we find
$  \Delta_r \equiv  \Delta_r' + i\Delta_r''
   \simeq \Delta + \frac{A}{\Delta}\log\left(1+\frac{\Delta^2}{\gamma_0^2}\right)
     \pm 2i\frac{A}{\Delta} \arctan\frac{\Delta}{\gamma_0}$,
where the real part $\Delta_r'$ is the renormalized frequency of the 
coherent oscillations, while the imaginary part $\Delta_r''$ describes
an exponential decay of those oscillations.  
If a Markovian approximation were made by setting $s=0$ in $\Gamma_1(s)$,
$\Gamma_y(s)$, and $K_y^+(s)$, then the bc would be missed completely 
and only an exponential decay with a rate $2A/\gamma_0$ would be obtained.
The Markov approximation is only justified if $\gamma_0\gg \Delta$, i.e., if
the bath dynamics is much faster than the system dynamics.
Here, we entirely avoid making a Markov approximation.

The Bromwich integral can then
be divided into two parts, $\langle\sigma_z(t)\rangle=
\langle\sigma_z(t)\rangle_{\rm poles}+\langle\sigma_z(t)\rangle_{\rm bc}$.
The integration in the first term  along the contour $C$, not including
the line integrals along the bc (Fig.~\ref{fig:analytic})
yields the sums of the residues from the poles
$\langle\sigma_z(t)\rangle_{\rm poles} 
        = \frac{1}{2\pi i}\int_C \! ds\,\langle\sigma_z(s)\rangle e^{st} 
        =  r' \cos(\Delta_r' t)e^{-\Delta_r'' t}
           - r'' \sin(\Delta_r' t)e^{-\Delta_r'' t}$,
where 
$r' = 1-(2A/\Delta^2)\log(1+\Delta^2/\gamma_0^2)+O(A^2)$
and
$r'' = (4 A/\Delta^2)\arctan(\Delta/\gamma_0) +O(A^2)$.
For $A=0$, this reduces to $\cos(\Delta t)$.

\begin{figure}[b]
\includegraphics[width=7.5cm]{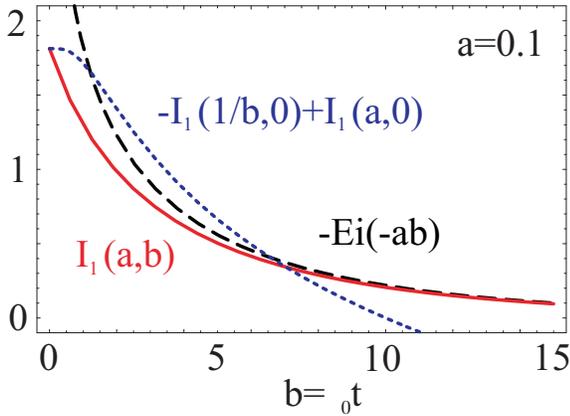}
\caption{(Color online) Branch cut integral function $I_1(a,b)$ (solid red line)
and two asymptotes (black dashed and blue dotted lines).
}
\label{fig:function}
\end{figure}

The branch-cut contribution to lowest order in $A$ is
\begin{equation}
 \langle\sigma_z(t)\rangle_{\rm bc}
   = \frac{4A}{\Delta^2} I_1(\gamma_0/\Delta,\Delta t)
\end{equation}
with the integral
$I_n(a,b)=\int_a^\infty dy \frac{e^{-by}}{y^n(y^2+1)^2}$,
where we have used Eq.~(\ref{eq:sol-trans-weak})
and introduced dimensionless variables and where $a>0$ and $b\ge 0$.
For $n=1$, we find (Fig.~\ref{fig:function})
\begin{eqnarray}
I_1(a,b) &=& \frac{1}{2}{\rm Re}\left[(ib+2)e^{-ib}(-i\pi+{\rm Ei}(ib-ab))\right]
\nonumber\\
& &-\frac{1}{2}\frac{1}{1+a^2}e^{-ab}-{\rm Ei}(-ab).
\end{eqnarray}
For $a=\gamma_0/\Delta > 1$ and $b>0$ ($t>0$), the 
effect of the environment from the bc integral is
exponentially suppressed: $I_1(a,b) < e^{-a b}/b$ and thus
$|\langle\sigma_z(t)\rangle_{\rm bc}|<(4A/\Delta^3 t) e^{-\gamma_0 t}$.
The physically more interesting regime is $a=\gamma_0/\Delta \ll 1$.
Within this regime, we can distinguish two temporal regimes:
short times $a b\ll 1$ ($t\ll\gamma_0^{-1}$)
and long times $a b \gg 1$ ($ t\gg \gamma_0^{-1}$).
In the short-time case, the integral is cut off from above by a 
combination of the $y^{-5}$ and the exponential factor.  The effect
of the latter can be approximated by cutting off the integral at $1/b$,
with the result $I_1(a,b)\approx -I_1(1/b,0)+I_1(a,0)$, where
$I_1 (a,0)=- \frac{1}{2}(1+a^2)^{-1}
+\frac{1}{4}\log(1+a^{-2})$ 
is the bc integral for $t=0$ ($b=0$).
Note that $I_1(a,0)\ge 0$ due to the logarithmic term.
In the long-time case, the integral is cut off by the exponential
whereas the $(y^2+1)^2$ factor in the denominator becomes irrelevant,
$I_1(a,b)\approx -{\rm Ei}(-ab)$.

At this point, the parameter that controls the strength of the 
non-Markovian effects due to 1/f noise can be identified as
$\xi=(A/\Delta^2)\log(1+\Delta^2/\gamma_0^2)$.
The regime of validity of the Born approximation
(the only approximation required in this paper) is confined
by the condition $\xi \ll 1$.  
The resulting damped qubit oscillation is plotted in
Fig.~\ref{fig:plot} for $A/\Delta^2=0.05$ and $\gamma_0/\Delta=0.05$
where $\xi \approx 0.1$.
If the infrared cutoff is lowered, the non-Markovian effects
due to 1/f noise become more pronounced.
However, since the dependence on the infrared cutoff $\gamma_0$ is 
only logarithmic, the result does not change drastically
even if $\gamma_0$ is much smaller than in our example, as
long as $A$ is chosen sufficiently small to ensure the validity of the
Born approximation.
E.g., for $\Delta\approx 10\,{\rm GHz}$ and $\gamma_0\approx 1\,{\rm
  Hz}$ (cf.\ Ref.~\onlinecite{Ithier05}) then
$\gamma_0/\Delta=10^{-10}$.  With \cite{fn_A} $A/\Delta^2=0.005$, one finds
a long-lived asymmetry as shown in the inset of Fig.~(\ref{fig:plot}).
The intermediate asymptotics of this contribution is
$\langle\sigma_z\rangle_{\rm bc}\approx\xi\approx 0.1$, 
while for longer times this contribution also decays logarithmically
to zero.  A similar long-time behavior has been found also for
longitudinal coupling \cite{Ithier05}.

\section{The biased case}
We again assume that the qubit prepared at time $t=0$ in one of the eigenstates 
$|0\rangle=|\!\!\uparrow\rangle$ of $\sigma_z$, i.e., 
$\langle {\bm \sigma}\rangle=(0,0,1)$, but now the qubit is biased,
$\epsilon\neq 0$.
In the absence of the fluctuators ($A=0$), the qubit would now undergo a
precession about an axis in the $xz$ plane with frequency $E/2\pi$,
where $E=\sqrt{\Delta^2+\epsilon^2}$.
In this unperturbed situation, $\langle\sigma_z(s)\rangle$ has three 
poles at $s=\pm iE$ and $s=0$, the former two giving rise to undamped
oscillations of  $\langle\sigma_z(t)\rangle$ with  frequency $E/2\pi$
and amplitude $\Delta^2/E^2$, while the latter allows for a 
non-vanishing stationary value $\epsilon^2/E^2$ of  $\langle\sigma_z(t)\rangle$ 
in the long-time limit.

\begin{figure}
\includegraphics[width=8.5cm]{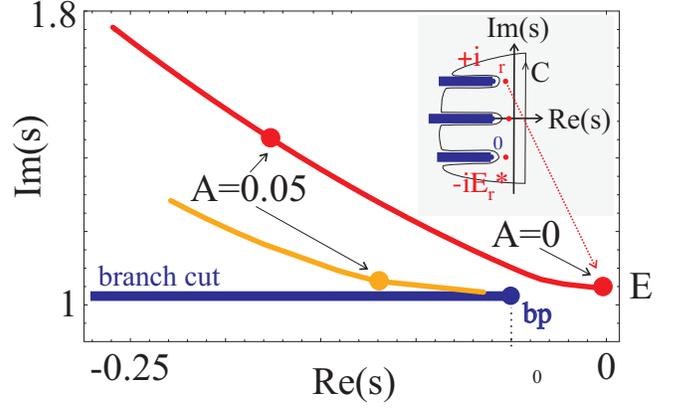}
\caption{(Color online) Shift and splitting of the poles of $\langle \sigma_z(s)\rangle$ 
for the biased system ($\epsilon =0.3$ and  $\gamma_0=0.05$).
Shown is the pole located at $s=iE$ for the undamped system ($A=0$),
indicated as a red dot (see Inset b).
The pole at $s=-iE$ behaves similarly.
With increasing  $A$ the pole shifts toward the vicinity of the
branch point (bp), where a second pole (orange dot) appears.
Shown as red and orange dots are the two poles for $A=0.05$.
The splitting of the poles leads
to a beating in $ \langle \sigma_z(t)\rangle_{\rm poles}$,
see Fig.~\ref{fig:plotbiased}.
Inset: Analytic structure of $\langle\sigma_z(s)\rangle$ for 
$\epsilon\neq 0$, where red dots are poles, and blue lines are
branch cuts (see Fig.~\ref{fig:analytic}).
}
\label{fig:poles}
\end{figure}

\begin{figure}
\includegraphics[width=8cm]{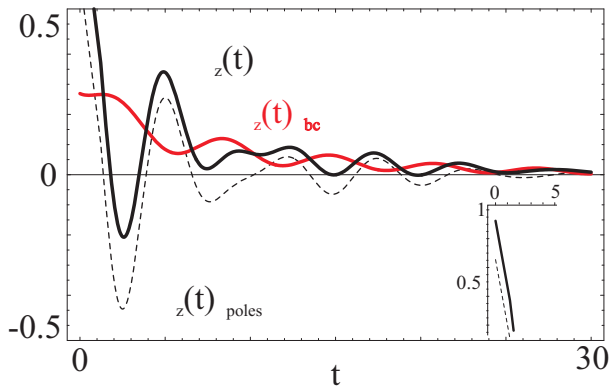}
\caption{(Color online)  Oscillation $\langle \sigma_z(t)\rangle$ of the biased qubit for $\epsilon/\Delta=0.3$, $A/\Delta^2=0.05$ and $\gamma_0/\Delta=0.05$. The beating due to the splitting of the poles at 
$\pm iE$ can be observed in $\langle \sigma_z(t)\rangle_{\rm poles}$.}
\label{fig:plotbiased}
\end{figure}

Including 1/f noise we find in leading order
in $A$ (see Appendix \ref{app:propagator}),
\begin{eqnarray}
  \langle \sigma_z(s)\rangle &=& 
\frac{s^2+\epsilon^2}{s(s^2+E^2)} 
+ 4 A\frac{\Delta^2}{E^2}{\rm Re}\left[
\frac{\Delta ^2}{\left(E^2+s^2\right)^2} C(s)\right.\nonumber\\
& & \left. + \frac{\epsilon ^2}{s^2 \left(s+iE\right)^2} C(s+iE)
\right]  +O(A^2) .
\end{eqnarray}
Analogously to the unbiased case, the poles are shifted in the
presence of the fluctuators. In leading order in $A$, we find three poles
at $-E_r'' = -(4 A \Delta^2/E^3) \arctan(E/\gamma_0)$, and
$\pm i E_r = \pm iE \pm (i A \Delta^2/E^3)\log(1+E^2/\gamma_0^2)
- (2 A \Delta^2/E^3)\arctan(E/\gamma_0)$.
From the shift of these poles (Fig.~\ref{fig:analytic}b), we obtain
$   \langle \sigma_z(t)\rangle_{\rm poles}
        = \frac{\Delta^2}{E^2}\cos(E_r' t)e^{-E_r'' t}
           +\frac{\epsilon^2}{E^2}e^{-2 E_r'' t}$.
However, while in the unbiased case a Markovian treatment at least
qualitatively describes the pole contribution correctly, in the biased case,
there is another effect that is elusive in a Markovian analysis.  
As shown in Fig.~\ref{fig:analytic}b, there are three bp's 
in the biased case, lying at $-\gamma_0$ and $-\gamma_0\pm iE$.
We find that as the two poles near $\pm iE$ approach the bp's
at $-\gamma_0\pm iE$ as $A$ is increased, these poles split into two
poles.  This behavior is illustrated in Fig.~\ref{fig:poles}.
The significance of this splitting is that it leads to beating patterns
already in the pole part of $\langle \sigma_z(t)\rangle$, as shown
in Fig.~\ref{fig:plotbiased}.
It should be noted that, again, the precise value of $\gamma_0$ is not critical
for the possibility to observe the effect, since $\gamma_0$ only
enters in the argument of a logarithm; even a much smaller value of 
$\gamma_0$  can thus be compensated by only a slight increase of
the system-environment coupling constant $A$.

The three bc's give
rise to a contribution to $\langle \sigma_z(t)\rangle$,
\begin{eqnarray}
  \langle \sigma_z(t)\rangle_{\rm bc}&=& -\frac{4 A \Delta^2}{E^4} 
\left[
\frac{\Delta^2+\epsilon^2\cos(Et)}{E^2} I_1\right.\nonumber\\
& &\left.+\frac{\epsilon^2}{E^2}\left(
 \sin(Et)  I_2 -\cos(Et)  I_3 \right)\right],
\end{eqnarray}
where the functions $I_n$ are as defined above and
are evaluated at the arguments $a=\gamma_0/E$ and $b=Et$.
For the unbiased case $\epsilon=0$ and $E=\Delta$, one
retrieves the previous result. The integrals $I_2$ and $I_3$ can
be calculated in closed form, but will not be given here.
The damped oscillations $\langle \sigma_z(t)\rangle$,
consisting of both pole and bc contributions, 
are plotted in Fig.~\ref{fig:plotbiased}.

\section{Comparison with an exactly solvable case}
The circumstance that in the case $\Delta=0$ the coupling Hamiltonian between the system
and the environment $H_{SB}$ commutes with the system Hamiltonian $H_S$ makes this
special case exactly solvable \cite{Martinis03,MS04,Rabenstein04,Schriefl06}.
A state prepared transverse to the common direction of the fixed precession axis and 
the fluctuating field, e.g., as $\langle {\bm\sigma}(t=0)\rangle=(1,0,0)$, for 
low-frequency noise essentially leads to a Gaussian decay behavior
$\langle \sigma_x(t)\rangle = \cos(\epsilon t)\exp\left(-ct^2\right)$.
%
The Born approximation which we have employed here can only be
expected to yield this result in lowest-order of the coupling constant, i.e.,
\begin{equation}\label{eq:Gaussiandecay}
\langle \sigma_x(t)\rangle \simeq \cos(\epsilon t)\left(1-ct^2+O(c^2t^4)\right).
\end{equation}
Here, we show that our result indeed has this form in the special case $\Delta=0$.

To this end, we take the limit $\Delta\rightarrow 0$ in the propagator, Eq.~(\ref{eq:GBE-L}),
as shown in the Appendix \ref{app:propagator}.  We then find 
\begin{equation}
\label{eq:sigmax}
\langle\sigma_x(s)\rangle = P_{xx}(s) 
 = \frac{s+\Gamma_y(s)}{\left(s+\Gamma_y(s)\right)^2+\left(\epsilon-\tilde{K}_y^+(s)\right)^2}.
\end{equation}
From Eq.~(\ref{eq:Cs}) and omitting logarithmic corrections, we can use $C(s)\simeq A/s$,
and thus $\Gamma_y(s)\simeq 4As/(s^2+\epsilon^2)$ and $\tilde{K}_y^+(s)\simeq 4A\epsilon/(s^2+\epsilon^2)$.  Substituting this into Eq.~(\ref{eq:sigmax}) and 
expanding to lowest order in $A$, we find
\begin{equation}
\langle\sigma_x(s)\rangle \simeq \frac{s}{s^2+\epsilon^2}+A\frac{s(3\epsilon^2-s^2)}{(s^2+\epsilon^2)^3},
\end{equation}
which equals the LT of Eq.~(\ref{eq:Gaussiandecay}) to lowest order, with
the identification $c=A/2$.  Therefore, our result is consistent with the
known exact result for $\Delta=0$, but, within the Born approximation, 
goes far beyond it, in that it 
includes arbitrary values of $\epsilon$ and $\Delta$.

\section{Discussion}
We find the following
essentially non-Markovian features in the decay of the $z$-component of the spin:
(i) The spin decay is non-exponential and asymmetric.
For relatively large infrared cutoff $\gamma_0$, 
there is an ``initial loss'' of coherence on a typical 
time scale $1/\gamma_0$, as seen in Figs.~\ref{fig:plot} and
\ref{fig:plotbiased}.  More importantly, for the typical case of small
$\gamma_0$, there is a long-time asymmetry favouring the qubit near its
initial state.
(ii) In the biased case, 1/f noise can lead to a two-frequency
oscillation, exhibiting a characteristic beating pattern.
Here, we have concentrated on the longitudinal component $\langle\sigma_z(t)\rangle$ 
of the qubit under the influence of both longitudinal and transverse 1/f noise.  
The transverse component $\langle\sigma_x(t)\rangle$ shows similar behavior.
The predicted non-Markovian effects are observable in
free induction decay (Ramsey fringe) experiments.
Indeed, such asymmetries are clearly visible in superconducting qubits 
\cite{Ithier05,Chiorescu03}
Measurements on a superconducting
flux qubit have shown deviations from the exponential decay and
beatings \cite{Deppe07}.  
The question whether these effects are due to the mechanisms described
here or not require further investigation.

\textit{Acknowledgments.}
Financial support for this work from the Swiss SNF (contract PP002-106310/1) 
and from German DFG SPP 1285 ``spintronics'' and FOR 912 is gratefully acknowledged.

\appendix

\section{Form of the propagator}
\label{app:propagator}
The propagator (resolvent) for solving the generalized Bloch
equation in Laplace space
is defined in Eq.~(\ref{eq:GBE-L}) as
\begin{equation}
P(s) = \left(s-R(s)\right)^{-1}.
\end{equation}
Using the form of the relaxation matrix $R(s)$,
we obtain the following expressions
for the matrix elements of $P(s)$,
\begin{eqnarray}
P_{xx}(s) &=& \frac{1}{D(s)}\left( s+ \Gamma_y(s) + \frac{\Delta ^2}{s} \right),\label{Pxx}\\
P_{yy}(s) &=& \frac{1}{D(s)}\left( s+ \frac{ E^2}{\Delta ^2} \Gamma_1(s)  \right),\label{Pyy}\\
P_{zz}(s) &=& \frac{1}{s}-\frac{\Delta^2}{s^2}P_{yy}(s) ,\label{Pzz}\\
P_{xy}(s) &=& -P_{yx}(s)=-\frac{1}{D(s)}\left(\epsilon-\frac{E}{\Delta }K_y^+(s)  \right),\label{Pxy}\\
P_{xz}(s) &=& P_{zx}(s) = -\frac{\Delta}{s} P_{xy}(s),\label{Pxz}\\
P_{yz}(s) &=& -P_{zy}(s) = -\frac{\Delta}{s} P_{yy}(s) ,\label{Pyz}
\end{eqnarray}
with the definition
\begin{eqnarray}
D(s) &=& \left(s+\Gamma_y(s)+\frac{\Delta ^2}{s}\right) 
          \left(s+\frac{E^2}{\Delta^2}\Gamma_1(s) \right) \nonumber\\
   & &   +\left(\epsilon -\frac{E}{\Delta }K_y^+(s) \right)^2.
\end{eqnarray}
The solution in Laplace space is now obtained according to Eq.~(\ref{eq:GBE-L}),
with ${\bf k}=0$,
\begin{equation}
\langle \sigma_i (s)\rangle = \sum_{j=x,y,z} P_{ij}(s) \langle \sigma_j(t=0)\rangle .
\end{equation}
E.g., for $\langle {\bm \sigma} (t=0)\rangle =(0,0,1)$, we find
$\langle \sigma_i (s)\rangle =P_{iz}(s)$.   Using Eqs.~(\ref{Pzz}), (\ref{Pxz}), and (\ref{Pyz}),
we recover the known results from Ref.~\onlinecite{DL05} 
in the special case ${\bf k}=0$.
The remaining matrix elements, Eqs.~(\ref{Pxx}), (\ref{Pyy}), and (\ref{Pxy}), 
allow us the use different initial conditions.

\subsection{The case $\epsilon=0$}
For an unbiased qubit,  $\epsilon=0$ and thus $E=\Delta$, 
so that the quantities discussed above are reduced to the form
\begin{eqnarray}
D(s) &=& \left(s+\Gamma_y(s)+\frac{\Delta^2}{s}\right)\left(s+\Gamma_1(s)\right),\\
P_{xx}(s)&=& \left(s+\Gamma_1(s)\right)^{-1},\\
P_{yy}(s)&=& \frac{s+\Gamma_1(s)}{D(s)}
     = \left(s+\Gamma_y(s)+\frac{\Delta^2}{s}\right)^{-1}\!\!\!\!,\\
P_{zz}(s)&=& (s+\Gamma_y(s))P_{yy}(s)/s,\\
P_{yz}(s)&=& -P_{zy}(s)=-\Delta P_{yy}(s)/s,\\
P_{xy}(s)&=&P_{yx}(s)=P_{xz}(s)=P_{zx}(s)=0.
\end{eqnarray}

\subsection{The case $\Delta=0$}
\label{app:prop-long}
In the case of a diagonal system Hamiltonian $H_S$, we set $\Delta=0$ and
thus $E=\epsilon$, and
\begin{eqnarray}
\Gamma_y(s) &=& \frac{E^2}{\Delta^2}\Gamma_1(s) = 
2A \left( C(s+i\epsilon) + C(s-i\epsilon)\right),\nonumber\\
 & & \\
\tilde{K}_y^+(s) &\equiv& \frac{E}{\Delta} K_y^+(s) = 2iA \left( C(s+i\epsilon) - C(s-i\epsilon)\right), \nonumber \\
& & \\
D(s) &=& \left(s+\Gamma_y(s)\right)^2+\left(\epsilon-\tilde{K}_y^+(s)\right)^2.
\end{eqnarray}
With Eqs.~(\ref{Pxx}--\ref{Pyz}), we obtain
\begin{eqnarray}
P_{xx}(s) &=& P_{yy}(s) = \frac{s+\Gamma_y(s)}{D(s)}, \\  
P_{zz}(s) &=& \frac{1}{s},\\
P_{xy}(s) &=& -P_{yx}(s) = -\frac{\epsilon-\tilde{K}_y^+(s)}{D(s)},\\
P_{xz}(s) &=& P_{zx}(s) = P_{yz}(s) = P_{zy}(s) = 0.
\end{eqnarray}

\end{document}